\documentclass[conference]{IEEEtran}
\IEEEoverridecommandlockouts
\usepackage{cite}
\usepackage{amsmath,amssymb,amsfonts}
\usepackage{algorithmic}
\usepackage{graphicx}
\usepackage{textcomp}
\usepackage{xcolor}
\usepackage{url}
\def\BibTeX{{\rm B\kern-.05em{\sc i\kern-.025em b}\kern-.08em
    T\kern-.1667em\lower.7ex\hbox{E}\kern-.125emX}}
\begin{document}

\makeatletter
\newcommand{\linebreakand}{%
    \end{@IEEEauthorhalign}
    \hfill\mbox{}\par
    \mbox{}\hfill\begin{@IEEEauthorhalign}
}
\makeatother

\newcommand{\ctext}[1]{\textcircled{\tiny#1}}

\title{Fabchain: Managing Audit-able 3D Print Job\\over Blockchain}

\author{
    \IEEEauthorblockN{Ryosuke Abe}
    \IEEEauthorblockA{\textit{Graduate School of}\\\textit{Media and Governance} \\
        \textit{Keio University}\\
        Kanagawa, Japan \\
        chike@sfc.wide.ad.jp}
    \and
    \IEEEauthorblockN{Shigeya Suzuki}
    \IEEEauthorblockA{\textit{Graduate School of}\\\textit{Media and Governance} \\
        \textit{Keio University}\\
        Kanagawa, Japan \\
        shigeya@sfc.wide.ad.jp}
    \and
    \IEEEauthorblockN{Kenji Saito}
    \IEEEauthorblockA{\textit{Graduate School of}\\\textit{Business and Finance} \\
        \textit{Waseda University}\\
        Tokyo, Japan\\
        ks91@aoni.waseda.jp}
    \linebreakand
    \IEEEauthorblockN{Hiroya Tanaka}
    \IEEEauthorblockA{\textit{Faculty of Environment}\\\textit{and Information Studies} \\
        \textit{Keio University}\\
        Kanagawa, Japan \\
        htanaka@sfc.keio.ac.jp}
    \and
    \IEEEauthorblockN{Osamu Nakamura}
    \IEEEauthorblockA{\textit{Faculty of Environment}\\\textit{and Information Studies} \\
        \textit{Keio University}\\
        Kanagawa, Japan \\
        osamu@sfc.wide.ad.jp}
    \and
    \IEEEauthorblockN{Jun Murai}
    \IEEEauthorblockA{\textit{Keio University} \\
        Tokyo Japan \\
        junsec@sfc.wide.ad.jp}
}

\maketitle

\begin{abstract}
    Improvements in fabrication devices such as 3D printers are becoming possible for personal fabrication to freely fabricate any products.
    To clarify who is liable for the product, the fabricator should keep the fabrication history in an immutable and sustainably accessible manner.
    In this paper, we propose a new scheme, ``Fabchain,'' that can record the fabrication history in such a manner.
    By utilizing a scheme that employs a blockchain as an audit-able communication channel, Fabchain manages print jobs for the fabricator's 3D printer over the blockchain, while maintaining a history of a print job.
    We implemented Fabchain on Ethereum and evaluated the performance for recording a print job.
    Our results demonstrate that Fabchain can complete communication of a print job sequence in less than 1 minute on the Ethereum test network.
    We conclude that Fabchain can manage a print job in a reasonable duration for 3D printing, while satisfying the requirements for immutability and sustainability.
\end{abstract}


\section{Introduction}
During the past decade, digital fabrication devices such as 3D printers have improved such that they can fabricate increasingly complex and delicate products.
As a result, many individuals now fabricate anything, possibly becoming to fabricate a house in the future~\cite{blikstein2013digital,gershenfeld2012make,digfab,SANJAYAN20191}.
This style of fabrication is called ``personal fabrication~\cite{gershenfeld2008fab,personalfab,chen2015direct}.''

As individual participation in personal fabrication increases, it is necessary to provide sustainable access to fabrication history to clarify who is liable for the product.
For example, if a 3D printed toy injures a child, the fabricator and the printed design data must be identified to ascertain the cause.
The Japanese Product Liability Act states that consumers can pursue compensation for damages from the fabricator of the product by proving that the design has a defect~\cite{JapanPL}.
Thus, it is necessary to record and maintain a history that identifies the fabricator, the design data, and the other fabrication environment data to ascertain the cause of the failure.

However, even if 3D printers record print job records as a type of history, the integrity of this history cannot be guaranteed without ensuring that no one has rewritten the history.
For example, even if a fabricator stored the history on a securely operated database, the database operator could still rewrite history.
Product liability acts in many countries define the period that the fabricator should be liable (e.g., in Japan, for ten years).
Thus, for recording and maintaining the history, it is a requirement that the history is immutable, and sustainably accessible while the fabricator is liable for products.

In this paper, we propose ``Fabchain,'' a new scheme that manages a 3D print job over a blockchain, thus recording 3D print job records as history on the blockchain.
Blockchain is a distributed ledger that records publicly immutable and highly available records in a peer-to-peer (P2P) network without a trusted third party.
Fabchain constructs the history in an immutable, and sustainably accessible manner by owing a scheme that uses a blockchain as an audit-able communication channel~\cite{bcaacc}.
We implemented a Fabchain prototype on the top of Ethereum, an application platform based on a blockchain.
To evaluate the performance of our implementation, we measured the time required for completing a print job.
As a security analysis, we discuss the security of Fabchain and compared it with a case using a public cloud service.

The remainder of this paper is organized as follows;
In Sec.~\ref{sec:background1}, we describe the fabrication process and issues related to recording history.
In Sec.~\ref{sec:blockchain}, we describe ``Blockchain,'' which is the key technology in this paper.
In Sec. \ref{Proposed}, we describe our proposed scheme ``Fabchain,'' and how the scheme satisfies the requirements of recording the history of print jobs.
In Sec.~\ref{Implementation}, we describe the details of our implementation of Fabchain.
In Sec.~\ref{Perfomance}, we evaluate the performance by measuring the time required to complete a print job.
In Sec.~\ref{Evaluation}, we discuss the security of our proposed scheme and compare it with a case using a public cloud service.
In Sec.~\ref{OpenProblems}, we discuss the open issues of Fabchain.
In Sec.~\ref{RelatedWorks}, we describe the related works of Fabchain.
In Sec.~\ref{Conclusion}, we conclude the paper.

\section{Personal Fabrication and Fabrication History}
\label{sec:background1}
First, we describe the model for personal fabrication processes that is our focus and the necessity of using history to clarify who is liable for product failures.
Second, we discuss the requirements for recording and maintaining the history.

\subsection{Fabrication Process of Personal Fabrication}
\label{sec:fabricationprocess}
Fig.~\ref{img:usecase} shows one of the use cases of personal fabrication.
There are three actors:
\begin{itemize}
    \item {\bf Fabricator}, which fabricates products with a 3D printer.
    \item {\bf E-store}, which lists and sells products fabricated by fabricators.
    \item {\bf Consumer}, who buys products.
\end{itemize}
Initially, the fabricator lists their product in the e-store.
At this time, the fabricator has not yet fabricated the product.
When the fabricator receives an order, the fabricator fabricates the ordered products on-demand.

\begin{figure}[h]
    \centering
    \includegraphics[width=\linewidth]{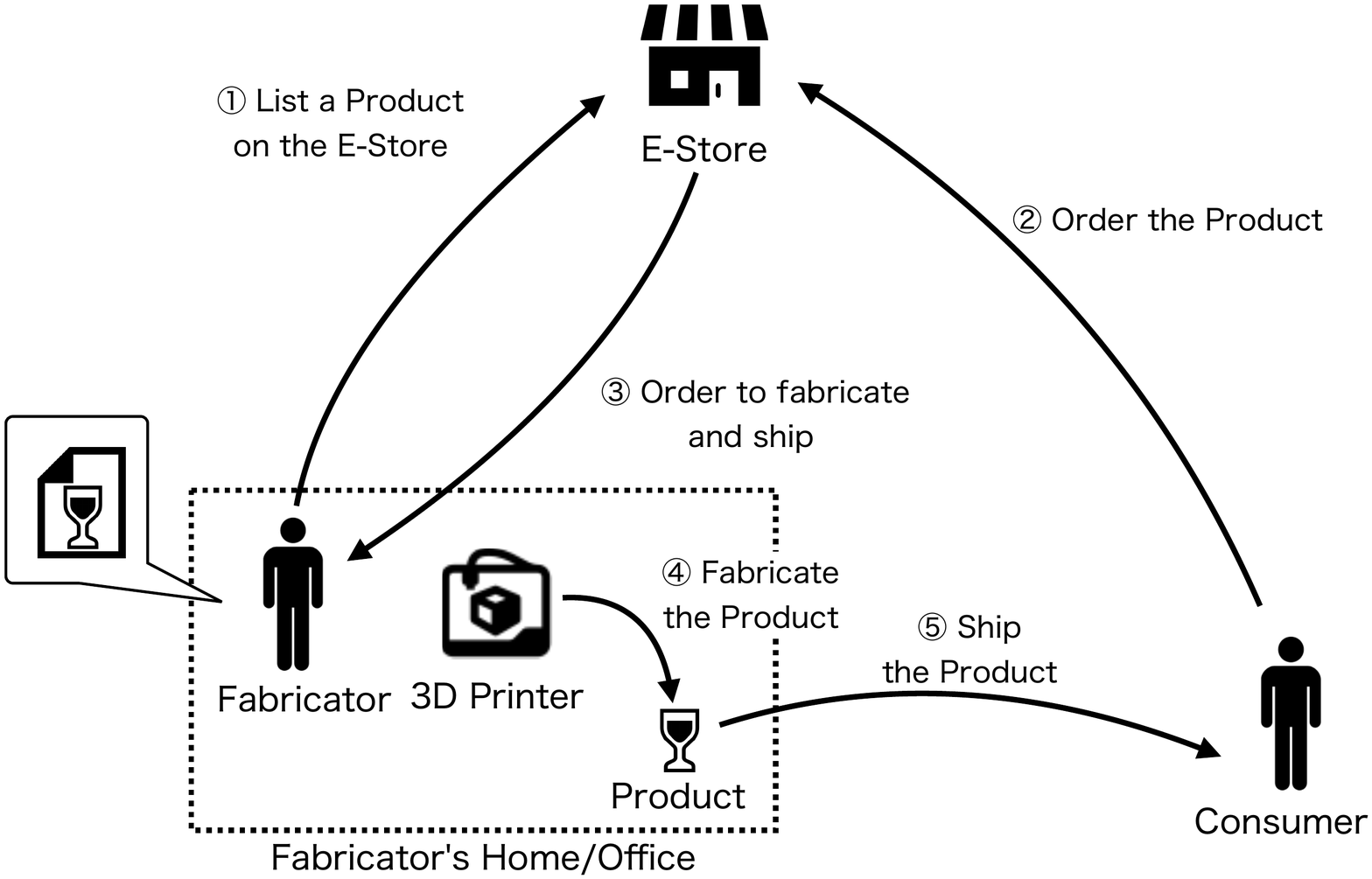}
    \caption{Use case of Personal Fabrication;
        First, the fabricator lists their product in the e-store (\ctext{1}).
        When the consumer orders the fabricator's product (\ctext{2}), the e-store requests the fabricator to fabricate and ship the product (\ctext{3}).
        The fabricator then fabricates the product with the fabricator's 3D printer (\ctext{4}) and ships it to the consumer (\ctext{5}}
    \label{img:usecase}
\end{figure}

In this use case, the fabrication history is created in the fabrication process.
In the process of fabrication, there are two actors:
\begin{itemize}
    \item {\bf Designer}, who creates a 3D model and publishes it on a 3D model publishing service.
    \item {\bf Fabricator}, who owns a 3D printer and fabricates a product.
\end{itemize}
The designer and the fabricator can be individuals, sole proprietorships, or small companies.
The designer and the fabricator own a PC to create and modify 3D models with Computer-Aided Design (CAD) software.
The designer can upload and publish 3D models that they have created to 3D model publishing services, such as Thingiverse\footnote{\url{https://www.thingiverse.com/}}.
The fabricator owns three devices:
\begin{itemize}
    \item {\bf 3D printer}, which prints a product from a 3D model.
    \item {\bf Print server}, which  manages print jobs and controls the 3D printer.
    \item {\bf Print client}, which is a PC for modifying 3D models and makes printing requests to the print server.
\end{itemize}
Fig.~\ref{img:process} shows one of typical fabrication processes in person fabrication.
In this process, the fabricator's devices (3D printer, print server, or print client) records the fabrication history in the devices' storage.

\begin{figure}[h]
    \centering
    \includegraphics[width=\linewidth]{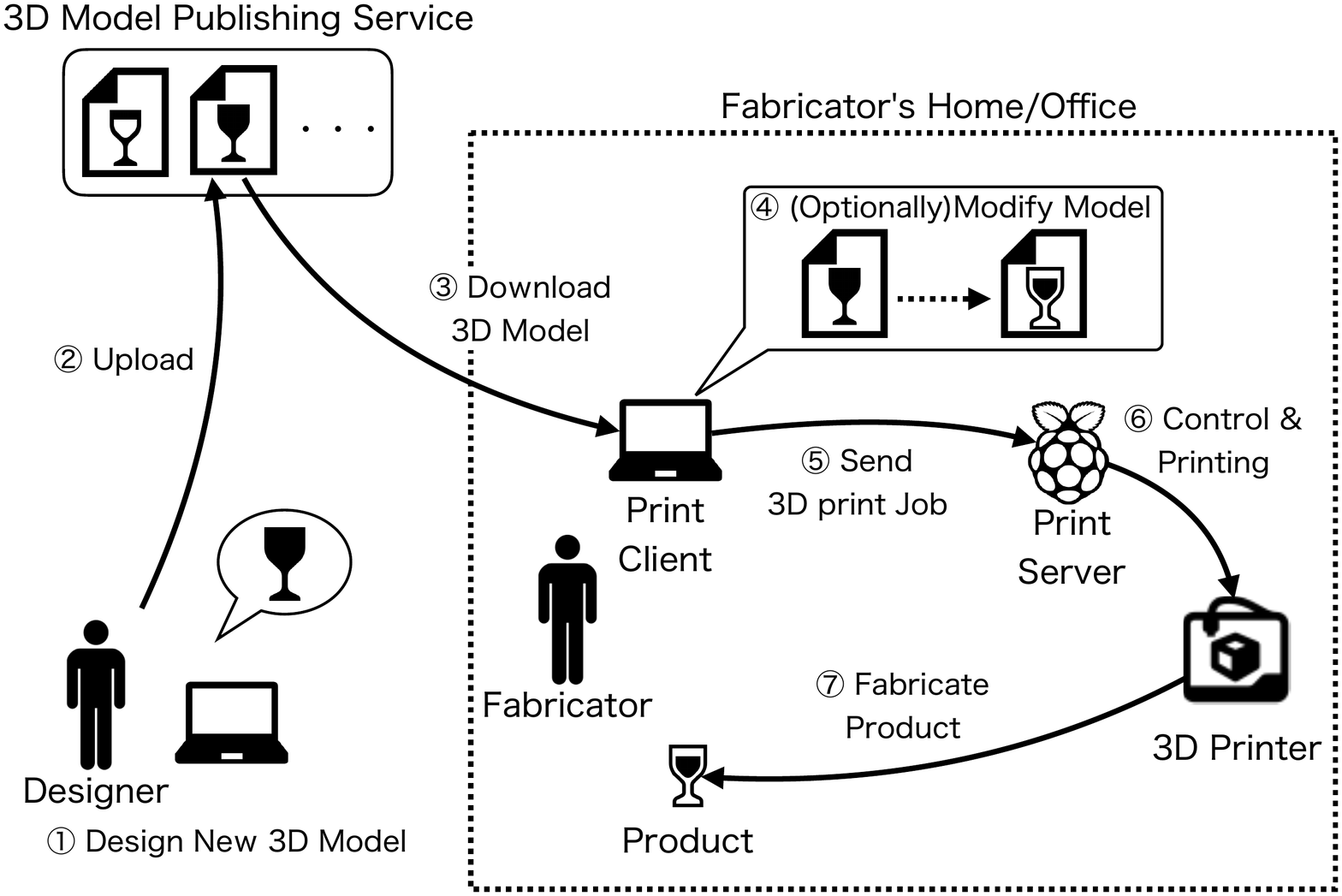}
    \caption{A Personal Fabrication Process with Optional Modification of the Design;
        First, a designer creates a new 3D model (\ctext{1}) and publishes it on a 3D model publishing service (\ctext{2}).
        A fabricator then downloads the 3D model on the client (\ctext{3}) and optionally modifies it to suit their use (\ctext{4}).
        The fabricator then sends a print job, including the 3D model, from the client to their print server (\ctext{5}).
        The print server controls the 3D printer (\ctext{6}) and fabricates a product from the 3D model (\ctext{7}).}
    \label{img:process}
\end{figure}

\subsection{Requirements to Clarify Liability}
\label{sec:requirement}
Clarifying liability requires a fabrication history that identifies the fabricator and the printed 3D model.
Hence, how to record and maintain that history is a significant issue.
Previous studies have employed various approaches to managing the fabrication history of products, including attaching a product serial identifier, embedding an RFID tag into the 3D printed product, or embedding an identifier as internal structure~\cite{fujiyoshi2014rfid,InfraStructs}.
In this paper, we focus on a scheme for recording and maintaining fabrication history specified by the product serial identifier.

To utilize fabrication history as evidence of product liability, the fabrication history must be {\it immutable} and {\it sustainably accessible}.
First, the history should be {\it immutable} to prevent anyone from rewriting said history, which would falsify claims of liability.
For example, a fabricator might delete the history of their product and then claim that they did not fabricate the product.
Alternatively, to fraudulently blame the fabricator or the designer, the consumer might falsify the history so that the consumer to pretend that the 3D model or the fabricating process has defects.

Next, the history should be {\it sustainably accessible}, that is, everyone should be able to verify the product liability.
When the consumer pursues product liability, the consumer should retrieve the history of the product and be able to verify whether the design has a defect.
According to the Japanese Product Liability Act, we estimate that the fabricator should keep history accessible for ten years~\cite{JapanPL}.

\section{Blockchain Technology}
\label{sec:blockchain}
In this section, we describe the ``Blockchain,'' which is the key technology in this paper.

\subsection{Blockchain Overview}
Blockchain technology is a distributed ledger that records publicly immutable records in a P2P network without a trusted third party.
In a blockchain system, a node records data in a unit called a ``Block.''
A block consists of the cryptographic hash value of the prior block and a series of units of data, called ``Transactions.''
Each node stores all the blocks in the node's local storage.

When a user records new data into the blockchain, the user creates a transaction and broadcasts it to the blockchain network.
Each node in the network verifies the transaction.
Some nodes create a new block from verified transactions and broadcast the new block.
In the block creation process, the nodes perform a cryptographic protocol called ``Proof-of-Work (PoW),'' wherein the nodes demonstrate to other nodes that they have performed a certain amount of computational work in a specified interval of time~\cite{PoW}.
Each node verifies the new block, and if verification succeeds, nodes store the block in local storage as a part of the blockchain.

When there are contradicting blocks, i.e., some blocks include the same hash value of the prior block with different transactions, each node chooses a block that spent the most computational power during the block creation process.
Because each block includes the previous block's hash value, if an attacker seeks to rewrite a historical block in a blockchain, the attacker will need computational power to create all blocks from the rewritten block to the latest block.
Because it is generally difficult to conduct such attacks, we can treat blockchain as an immutable ledger.

Because a node with the entire blockchain can independently verify blocks and transactions, the blockchain achieves an immutable ledger without a trusted third party.
Blockchains are also highly available because all nodes in the blockchain network replicate and maintain the blockchain.
Those characteristics of the blockchain are considered useful in applications, in addition to cryptocurrency~\cite{7573685,khan2018iot,namecoin}.

\subsection{Ethereum}
Ethereum is a blockchain-based application platform on which users can run programs called ``smart contracts~\cite{ethereumyellow}.''
An Ethereum node has the blockchain and an Ethereum Virtual Machine (EVM), which includes an interpreter and the World State consisting of contract states for each smart contract.
When the node records a transaction that includes a smart contract into the blockchain, the EVM creates a contract state and executes the constructor of the smart contract (we refer to this condition as ``a smart contract is deployed.'').
The transaction includes the digital signature of the issuer of the transaction.
The issuer is identified by the ``Ethereum Address,'' which is the hash value of its public key.
After creating the contract state, the smart contract is identified by the smart contract pointer called the ``Contract Address.''

To update the contract state, a user broadcasts a transaction that includes the contract address and input values for a function in the smart contract.
When the node records a block with the transaction into the blockchain, the EVM executes the function and updates the contract state.
With this scheme, because the node records each transition of the contract state in an immutable and highly available manner, the contract state is also immutable and highly available.
When simply reading the contract state, no transaction is required.

With Ethereum, a transaction for deploying and executing a smart contract requires execution fees.
The fee amount for the EVM opcodes is defined in the Ethereum Yellow paper and is hardcoded in the EVM implementation~\cite{ethereumyellow}.

\subsection{Blockchain as Audit-able Communication Channel}
\label{sec:baacc}
Suzuki and Murai proposed to use blockchain as an audit-able communication channel~\cite{bcaacc}.
In their scheme, a client creates a transaction that includes a message to a server.
The server retrieves the transaction from the blockchain and obtains the message.
With the scheme, the history of the communication is recorded in an immutable, highly available, and verifiable manner.
Past studies utilized this scheme in IoT (Internet of things) applications and human-robot interactions~\cite{haddadi2018siotome,ferrer2018robochain}.

\section{Fabchain: Managing Audit-able 3D Print Job over Blockchain}
\label{Proposed}
In this paper, by utilizing the scheme described in Sec.~\ref{sec:baacc}, we propose ``Fabchain,'' a scheme that manages 3D print jobs over a blockchain.

\subsection{Proposed Scheme}
Fig.~\ref{img:Fabchain} shows the overview of Fabchain.
A fabricator posts a request transaction that includes a 3D printing request (\ctext{1} and \ctext{2}).
After recording the request transaction into the blockchain, a print server that controls a 3D printer retrieves the transaction from the blockchain and obtains the request (\ctext{3}).
After printing (\ctext{4}), the print server posts a response transaction that includes fabricating results (\ctext{5}).
With the scheme, the status and data of the print job are managed over the blockchain.
For example, if there is only the request on the blockchain, the print job has not been completed.

\begin{figure}[h]
    \centering
    \includegraphics[width=\linewidth]{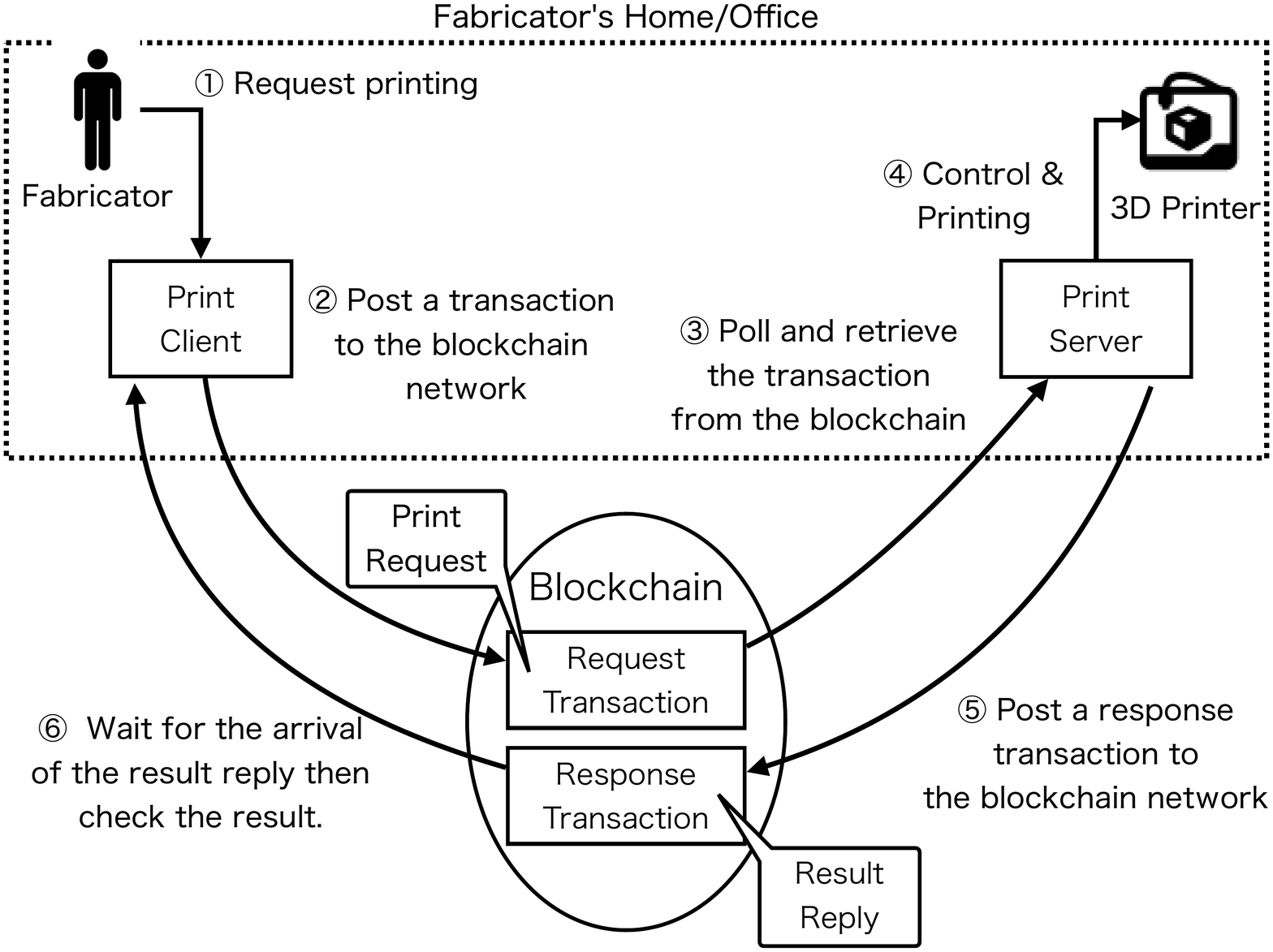}
    \caption{Fabchain Overview}
    \label{img:Fabchain}
\end{figure}

\subsection{Satisfying Requirements}
In this section, we describe how Fabchain satisfies each requirement mentioned in Sec.~\ref{sec:requirement}.
In this paper, we discuss Fabchain as implemented on top of Ethereum.

\subsubsection{Immutable}
The fabrication history within Fabchain is immutable according to the immutability of the blockchain.
The immutability of blockchain is described in the Bitcoin white paper~\cite{nakamoto2008bitcoin}.
It can be adopted to Ethereum because Ethereum also has immutability based on the PoW.
The white paper describes that the more blocks that are stacked on the block that is the target of rewriting, the more immutable is the block.
In Ethereum, the interval for the block creation is approximately 12 seconds.
When fabricating with a 3D printer, the fabrication process will typically takes a few hour.
Therefore, while fabricating, the history would become immutable by stacking some blocks on the block that includes the request transaction.

\subsubsection{Sustainably Accessible}
With Fabchain, the history is public because the Ethereum main network is a public network.
We can obtain the history on the blockchain by accessing the blockchain from any online Ethereum nodes.
Therefore, history is accessible as long as one of the nodes remains in the Ethereum network.

\section{Implementation}
\label{Implementation}
We implemented a proof-of-concept version of Fabchain on top of Ethereum.
We defined the objective of the implementation as the ability to identify the printed 3D model.
Thus, we implemented Fabchain so that it records the hash value of a printed 3D model.
Methods to trace the original designer of the printed 3D model are out-of-scope for the current implementation.

This implementation consists of the followings components:
\begin{itemize}
    \item {\bf Print client}, which creates a request transaction that includes a print request.
    \item {\bf Smart contract}, which manages the status of print jobs and provides interfaces for accessing history in the contract state.
    \item {\bf Print server}, which retrieves print requests, creates a response transaction that includes the results of the print request, and controls a 3D printer.
\end{itemize}
We implemented the print client and the print server as command-line applications in Python, and the smart contract in Solidity~\cite{soliditydoc}.
Fig.~\ref{img:implementation} shows an overview of the processing of a print job in our implementation.
The client and the server interact with the smart contract via possibly different Ethereum nodes.
The Ethereum nodes participate in an Ethereum network that the smart contract has been deployed.
The client and server have corresponding private keys that the fabricator generated in advance.
The client and the server are identified by their Ethereum Addresses.

\begin{figure}[h]
    \centering
    \includegraphics[width=\linewidth]{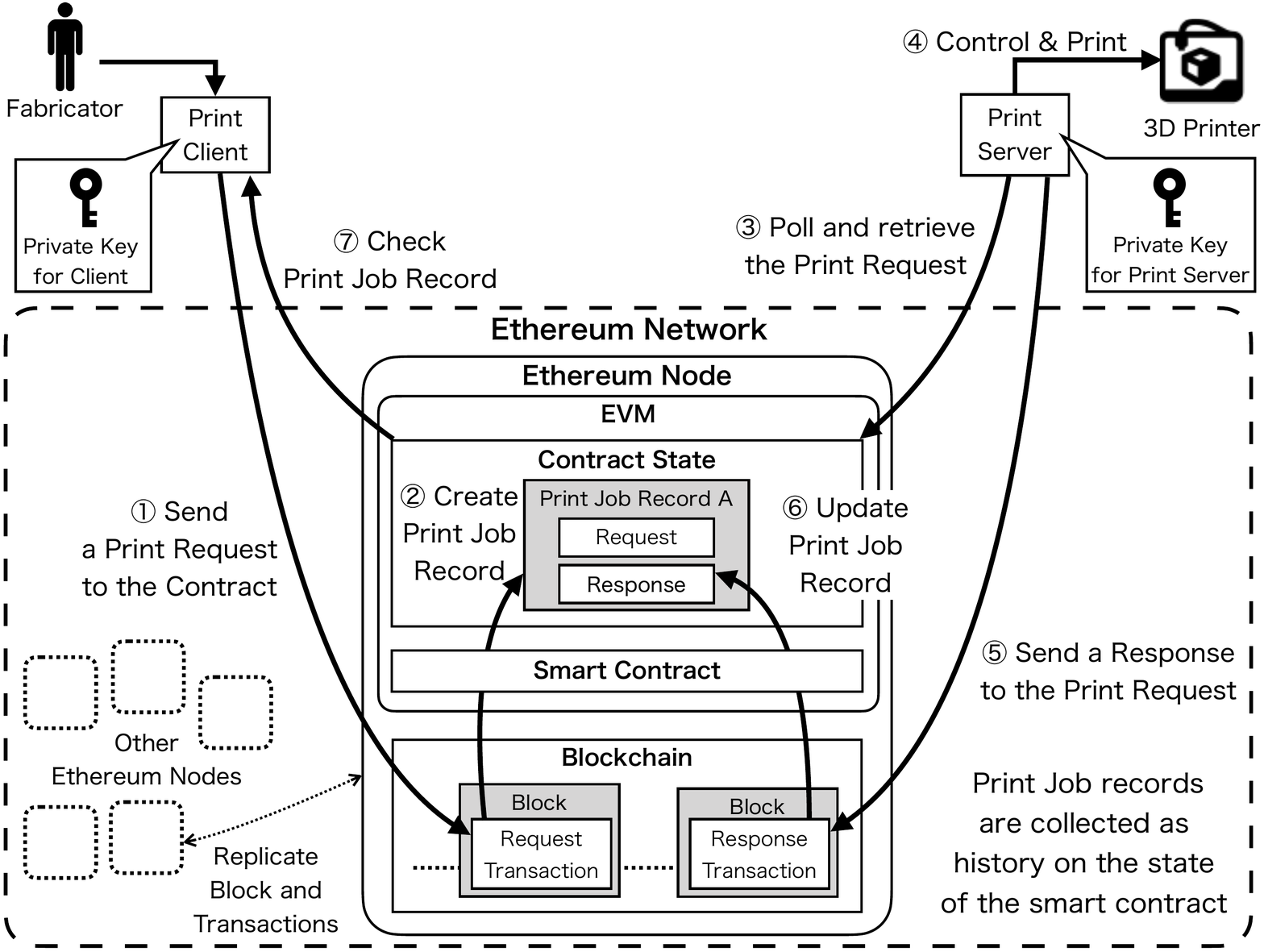}
    \caption{Implementation Overview; This figure shows the process for the product fabricated by Print Job A.
        The smart contract collects historical print job records as a history and provides that history so that anyone can read.}
    \label{img:implementation}
\end{figure}

The overview of the process in our implementation is as follows.
When the client sends a request transaction (\ctext{1}), the smart contract creates a new print job record (\ctext{2}).
The print job record is identified by the Print Job ID, which is the request's hash.
Anyone then can retrieve the print job records using the Print Job ID.
The server polls the contract state to retrieve new print job records (\ctext{3}).
When the server finds a new print job record, the server verifies and adds the request to the print queue.
The server waits for the print to finish (\ctext{4}), then sends a response transaction (\ctext{5}).
The contract then records the response to the print job record (\ctext{6}).
During the entire process, the client waits for the update of the print job record by polling.
The client finally receives the result in the print job record and does post-print processes accordingly (\ctext{7}).

\subsection{Assumptions}
We state the following assumptions for our implementation.
First, we assume that consumers can obtain a Print Job ID as a serial identifier that identifies the product by technology-agnostic schemes.
For example, each product may contain a serial identifier embedded in an RFID tag.

Second, we assume that entities use Ethereum Addresses as identifiers of each entity, however, how to authenticate the owner of the Ethereum Address is out of the scope of the paper.

Third, we assume to use an external data repository that stores data that is difficult to store on the Ethereum blockchain.
In Ethereum, it is difficult to manage large data, such as 3D model data, due to the size limitations of the data type of Solidity~\cite{soliditydoc}.
It is a typical scheme that stores large data as a value on a Key-Value Store (KVS) and stores the key to a blockchain~\cite{8946164,9027313}.
By using a hash value of the data as the key, the user can verify whether the retrieved data is the data that is identified by that key.
In this paper, we assume that the data is highly available by storing on a highly available KVS such as IPFS~\footnote{\url{https://ipfs.io/}}.
To simplify our implementation, we implemented a simple KVS to store and retrieve data using the hash value and utilized this implementation~\footnote{\url{https://github.com/chike0905/simple-hashstorage}}.

\subsection{Data Structures}
\label{sec:impl:data}
In this section, we describe the data structures used in the contract state: Print Job record, and Request record.
The fields in the Print Job record are shown in Table~\ref{table:log}.
Print Job record includes the Request record, the printing date, and flags that show the condition of completion of approval and print.

\begin{table}[t]
    \caption{Fields in Print Job Record}
    \label{table:log}
    \centering
    \begin{tabular}{|c|c|c|}
        \hline
        Parameter & \begin{tabular}{c}Type\end{tabular} & Description                \\
        \hline
        \hline
        Request   & Request                    & \begin{tabular}{c}Fabrication Request\\described in Table~\ref{table:request}\end{tabular} \\
        \hline
        PrintDate & bytes32                    & \begin{tabular}{c}Timestamp\\when the response transaction\\is generated\end{tabular} \\
        \hline
        Approved  & bool                       & \begin{tabular}{c}Request is approved or\\not approved by the server\end{tabular} \\
        \hline
        Printed   & bool                       & \begin{tabular}{c}The print job is printed or\\not printed by the server\end{tabular} \\
        \hline
    \end{tabular}
\end{table}

The fields in the Request record are shown in Table~\ref{table:request}.
The Request record includes the hash values of the 3D model to retrieve those data from the data repository.

\begin{table}[t]
    \caption{Fields in Request Record}
    \label{table:request}
    \centering
    \begin{tabular}{|c|c|c|}
        \hline
        Parameter & \begin{tabular}{c}Type\end{tabular} & Description                    \\
        \hline
        \hline
        From      & Address                    & Ethereum Address of fabricator \\
        \hline
        Printer   & Address                    & Ethereum Address of 3D printer \\
        \hline
        ModelHash & Bytes32                    & \begin{tabular}{c}Hash value of\\requested 3D model data\end{tabular}     \\
        \hline
        Date      & Bytes32                    & \begin{tabular}{c}Timestamp\\when the request is generated\end{tabular}     \\
        \hline
    \end{tabular}
\end{table}

\subsection{Print Job Sequence}
\label{sec:impl:process}

\begin{figure*}[t]
    \centering
    \includegraphics[width=\linewidth]{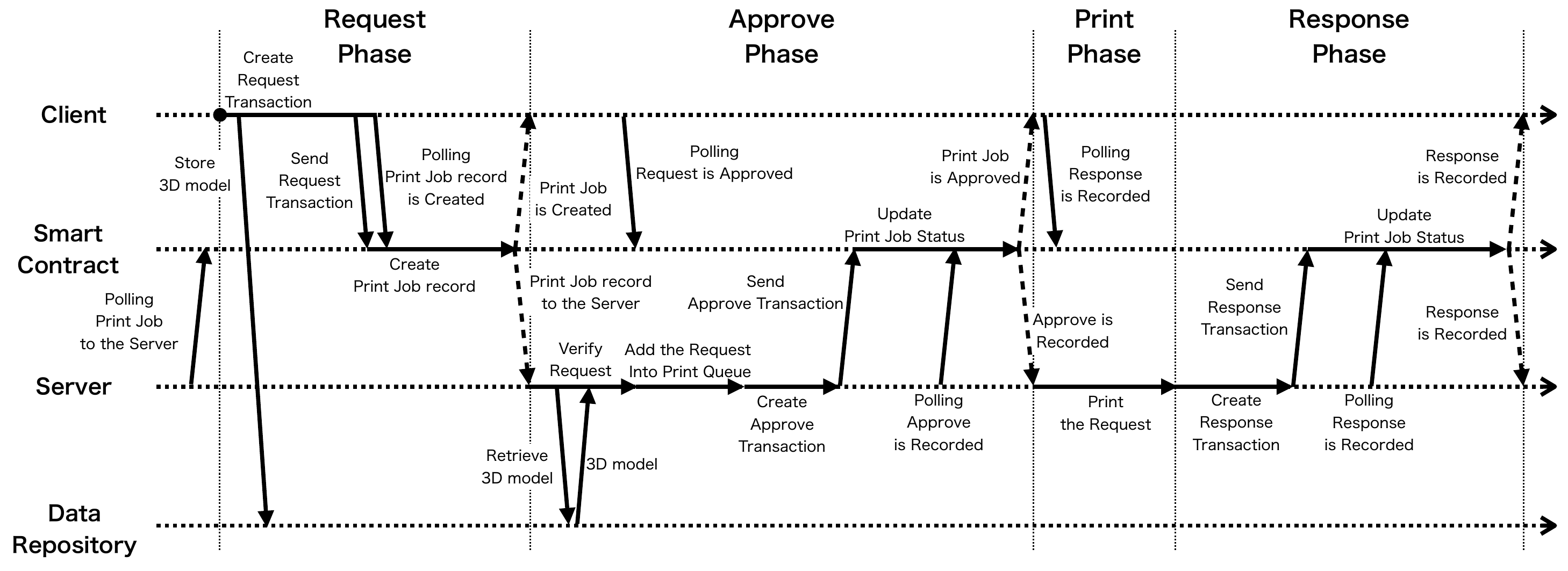}
    \caption{Sequence of the Print Job Process}
    \label{img:impl}
\end{figure*}

In this section, we describe the print job sequence.
Table~\ref{table:log} and Table~\ref{table:request} show the data structures in the contract state: the Print Job record and the Request record.
Fig.~\ref{img:impl} shows the sequence of the process.
The process is separated into four phases:
\begin{enumerate}
    \item {\bf Request Phase:} the client creates a new Print Job record on the contract state.
    \item {\bf Approve Phase:} the server approves the Request in the Print Job record.
    \item {\bf Print Phase:} the server prints the print request using the connected 3D printer.
    \item {\bf Response Phase:} the server responds to the Request in the Print Job record.
\end{enumerate}
The client performs only the Request phase and, after the request phase, keeps polling the print job status in the contract state.
The server keeps polling the contract state to find a new Print Job to the server.

\section{Performance Analysis}
\label{Perfomance}
We evaluated the performance of Fabchain by measuring the time required for each phase described in Sec.~\ref{sec:impl:process}.
We experimented on an Ethereum test network.
We then estimated the performance on the main network with parameters in a previous study, which measured the performance of a smart contract~\cite{ethlatency}.

\subsection{Estimation of Performance}
Before the experiment, we clarified parameters that affect the performance and estimate the performance.
In this paper, we defined the duration for a transaction calling a contract function as $D_{tx}$.
In Ethereum, $D_\mathrm{tx}$ depends on the duration from when the transaction is submitted to when the transaction is recorded in the blockchain.
Hence, the $D_\mathrm{tx}$ is defined by the following parameters:
\begin{itemize}
    \item $D_\mathrm{block}$: The duration between each block creation event
    \item $N_\mathrm{UntilIncluded}$: The number of blocks between the latest block when a transaction is submitted and the block that includes the transaction.
\end{itemize}
Because the elapsed time since the last block is determined when a transaction is submitted, we can estimate the value of $D_\mathrm{tx}$ would be as follows.
\begin{eqnarray}
    \label{eq:txinterval}
    \begin{split}
        &D_\mathrm{tx} \leq D_\mathrm{block} \cdot N_\mathrm{UntilIncluded}
    \end{split}
\end{eqnarray}
We estimated the duration for a phase that includes one transaction process, as defined by (\ref{eq:txinterval}).

\subsection{Experimental Setup}
For the experiment, we prepared four virtual machines as a client, a server, a data repository, and an Ethereum node.
For the experimental setup, we connected the client and the server connects to the same Ethereum node because the client and the server are in the fabricator's home according to the use case mentioned in Sec.~\ref{sec:fabricationprocess}.
Table~\ref{table:experimentenviroment} shows the configuration of the four virtual machines.
We used an Ethereum node implementation known as ``Geth~\cite{geth}.''
We experimented with our implementation on a node that participates in an Ethereum test network called ``Ropsten.''
The configuration for Ropsten is the same as the main network:
the network is public, and $D_\mathrm{block}$ is adjusted as an average of 12 seconds.
The difference between the Ropsten network and the main network is that the cryptocurrency on Ropsten has no value, and participants of Ropsten are fewer than in the main network.
Hence, we can experiment on Ropsten as an emulation of the main network.
The client and the server communicate with the Geth node via the RPC API.
We processed the experiment remotely via the console and obtained processing logs on each of the virtual machines.
We measured the print job sequences 100 times.

\begin{table}[t]
    \caption{Configuration of Virtual Machines}
    \label{table:experimentenviroment}
    \centering
    \begin{tabular}{|c|c|}
        \hline
        Item                & Configuration                          \\
        \hline
        \hline
        Operating System    & Debian 10                              \\
        \hline
        VM Config           & 4GB memory, 4 vCPUs                    \\
        \hline
        Hypervisor          & ESXi 6.5                               \\
        \hline
        Hypervisor Hardware & Fujitsu rx200 s6 (48GB Memory, 24CPUs) \\
        \hline
    \end{tabular}
\end{table}

\subsection{Results}
First, we analyzed the log of Geth to verify $D_\mathrm{tx}$.
Fig.~\ref{img:txintarval-ropsten} shows the experimental results.
In the experiment on Ropsten, it is difficult to measure when a node in Ropsten creates a block.
Thus, we instead take the time when our node receives a block as the time when the block is created.
In the experiment on Ropsten, $D_\mathrm{block}$ was 9.17 seconds, $D_\mathrm{tx}$ is was 15.90 seconds, and $N_\mathrm{UntilIncluded}$ was 1 blocks on average.
The result was slightly different from our estimation because there was overheads for propagating transactions and blocks from nodes.

\begin{figure}[h]
    \centering
    \includegraphics[width=\linewidth]{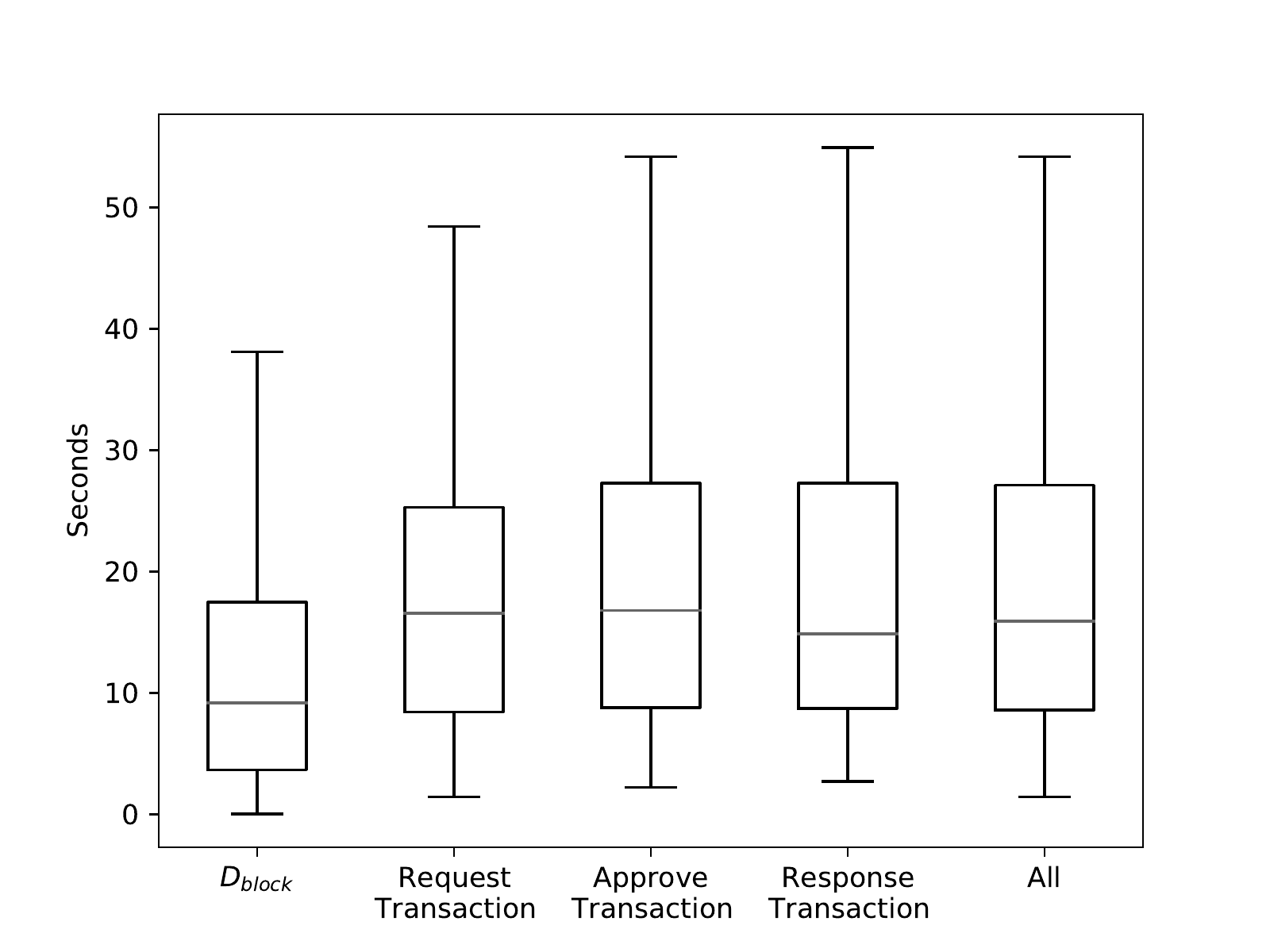}
    \caption{Duration from submitting transaction}
    \label{img:txintarval-ropsten}
\end{figure}

Next, we measured the duration of each phase.
We measured the request phase from logs on the client and other phases from logs on the server.
We also measured the duration from the client starts the request phase to the client can detect the request is approved (``request-approve''), and from the client starts to the client detects the response is recorded (``All Phase'').
In the measurement, we eliminated the print phase because the duration for printing is affected by the 3D printer and 3D model.

Fig.~\ref{img:request-response-ropsten} shows the results.
The resulting duration of request-approve was approximately 38.93 seconds, and the result of the request phase was approximately 14.66 seconds.
The distribution of those results were in accord with the distribution of $D_\mathrm{tx}$.
Thus, we concluded that our estimation is reasonable on Ropsten.

\begin{figure}[h]
    \centering
    \includegraphics[width=\linewidth]{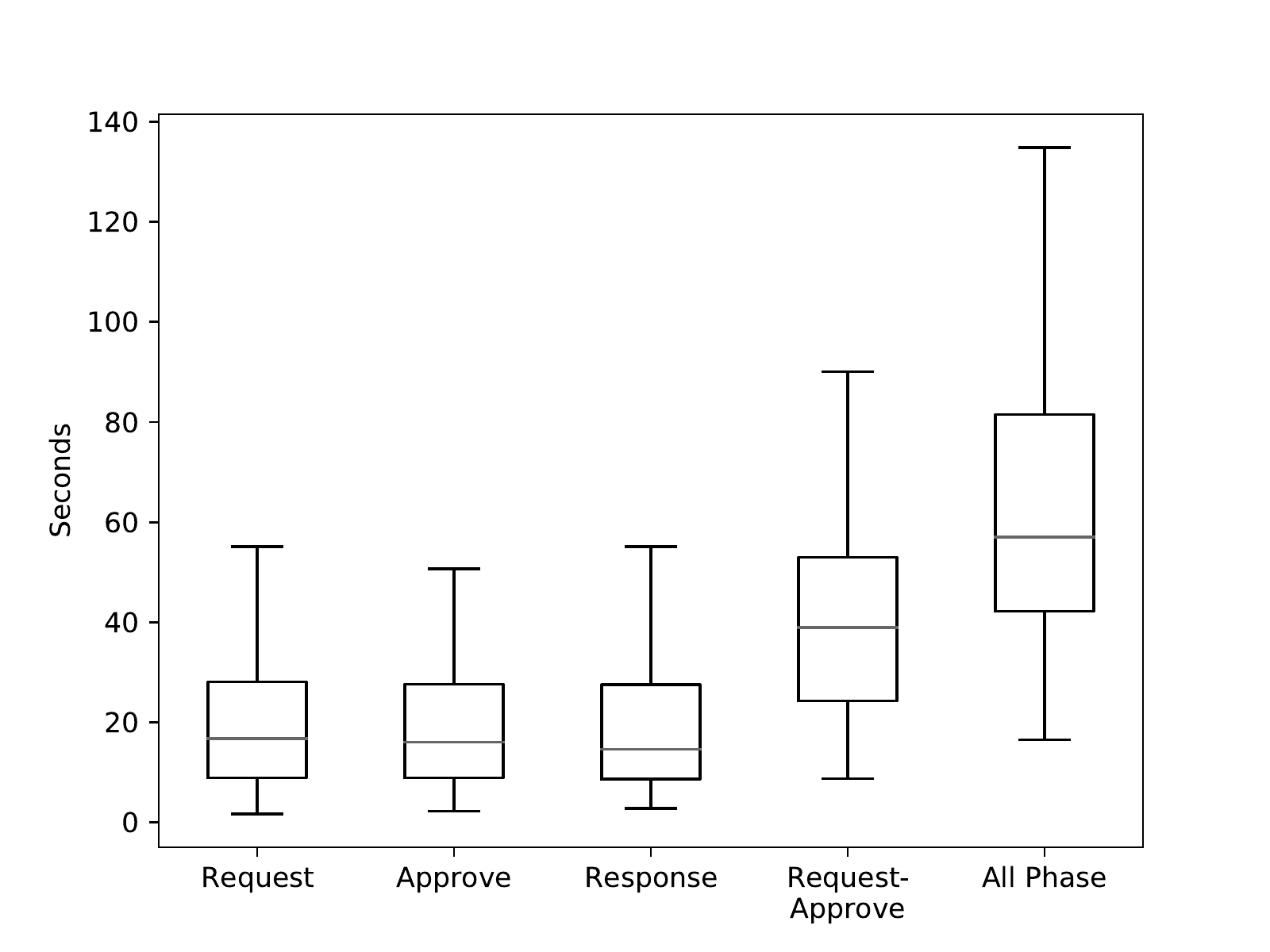}
    \caption{Duration of Each Phase}
    \label{img:request-response-ropsten}
\end{figure}

\subsection{Discussion: Estimation on Ethereum Main Network}
In the Ethereum main network, the $D_\mathrm{block}$ is 12 seconds\footnote{\url{https://etherscan.io/chart/blocktime}}.
Spain et al. measured the latency of transactions on the Ethereum main network~\cite{ethlatency}.
They concluded that over half of all transactions issued at depth $i$ in the blockchain were included by the block at depth $i + 2$.
Therefore, we can estimate the $N_{UntilIncluded}$ is approximately two blocks from their result.
With those parameters, the duration for request-approved would be 48 seconds.
The duration for the response phase would be 24 seconds.
Even if we consider the overhead for the propagation of transactions and blocks, we concluded that this duration does not present a problem because 3D printing takes significantly more time, even when printing small objects.

\section{Security Analysis}
\label{Evaluation}
In this section, we compare Fabchain and a case in which a fabricator constructs a history publishing service, from a security perspective.
Namely, we address how to satisfy the requirements mentioned in Sec.~\ref{sec:requirement}.

For comparing each scheme, we defined three setups to operate Fabchain or a history publishing service.
\begin{itemize}
    \item {\bf Setup 1} A fabricator operates the client and the server of Fabchain and an Ethereum node as a gateway to the Ethereum network.
    \item {\bf Setup 2} A fabricator operates the client and the server of Fabchain and utilizes an Ethereum node that other entities, such as Infura operate~\footnote{\url{https://infura.io/}}.
    \item {\bf Setup 3} A fabricator operates a history publishing service on a cloud computing service, such as Google Cloud Platform~\cite{GCP}.
\end{itemize}

\subsubsection{Security of Fabchain}
In Fabchain, the immutability of the history is provided by the underlying blockchain.
If one of the operators of the Ethereum nodes tries to fake history on a blockchain in their node, other nodes do not accept the faked history.
The fabricator and consumers do not need to trust any operator of a specific Ethereum node.

From the perspective of sustainable accessibility, history by Fabchain is highly available.
In December 2021, several entities are operating approximately 2000 Ethereum nodes all over the world~\footnote{\url{https://etherscan.io/nodetracker}}.
It is not realistic that all the nodes suddenly leave the network.
Hence, history would be sustainably accessible semi-permanently.

\subsubsection{Security of history publishing service}
In the case of a history publishing service, it is difficult to prove that the history is immutable.
For example, when the fabricator is malicious and makes it so that the history publishing service provides a faked history, the consumer cannot verify whether the provided history is faked or not because there is no proof.

In another aspect, the availability of the history publishing service depends on the operator of the instances on the public computing service.
To keep history sustainably accessible, a specific operator should keep operating the history publishing service.
When the fabricator operates the history publishing service for its products, the history is lost if the fabricator stops working as a fabricator and operating the instance.

\subsubsection{Security Discussion on Each Setup}
In Setup 1, the fabricator's Ethereum node broadcasts transactions of Fabchain to other Ethereum network participants.
The fabricator's Ethereum node can independently verify whether the transaction is recorded in the blockchain.
Hence, Setup 1 can satisfy the requirements securely.

In Setup 2, there is a risk that the gateway node is malicious.
For example, the gateway node can deceive whether the transaction is recorded.
However, there are several lightweight node schemes, such as Simple Payment Verification, that can verify whether a transaction is included in the blockchain~\cite{SPV}.
Combining those lightweight node schemes, Setup 2 can also satisfy the requirements while mitigating the risk.

In Setup 3, it is difficult to satisfy the requirements because there is no proof of immutability.
Considering the operation of the history publishing service, we conclude there is an advantage for Fabchain.

\section{Open Issues}
\label{OpenProblems}
In this section, we describe three open issues in Fabchain.
The first issue involves managing the identity of each entity.
The second issue involves the volatility of the cost for Fabchain.
The third pertains to the applicability of Fabchain.

\subsection{Identity Management}
In our implementation, we utilized an Ethereum Address as an identifier of a fabricator.
To pursue a product's liability, it is necessary to identify the fabricator from the print job record.
Therefore, it is a requirements of future research for Fabchain to develop methods to identify the owner of Ethereum Address.
One method would be to embed the public key certificate into the Print Job record.
By verifying the public key certificate and the Ethereum Address, the method can verify that the Print Job record is issued with the private key owned by the entity described in the public key certificate.

\subsection{Volatility of Cost}
Because the price of ether is highly volatile, the cost for a print job with Fabchain may be high.
While working on this research from 2020 March to 2021 December, the cost for Fabchain became 20,000 times higher than when we first estimated the cost.
The issue is caused by the design of Ethereum, wherein a user pays a fee for executing a smart contract.
Blockchain-based applications that do not relate to cryptocurrency share this issue~\cite{8605977}.
Re-designing a blockchain as an application platform without cryptocurrency is one possible solution for this problem.

\subsection{Applicability of Fabchain}
In this section, we discuss the applicability of Fabchain.
First, fabrication history is useful for evaluating fabrication processes.
Various factors, such as temperature and 3D model structures, can cause digital fabrication devices to fail when creating products.
Thus, the detailed results are important to analyze the reasons for the failure.
Hence, we need to determine what data should be recorded in the history.

\section{Related Works}
\label{RelatedWorks}
Several studies have applied blockchain technology to tasks in industries~\cite{industrialbc,dedeoglu2019journey}.
With the widening adoption of ``Industry 4.0,'' which involves collaborating with several parties in the industrial process via the Internet, previous studies have applied blockchain to guarantee trust within parties~\cite{bcin40,8256072,angrish2018case,schaffers2018relevance}.
Makerchain, one of the previous studies, is a blockchain-based system that records multi-party activity in Industry 4.0~\cite{leng2019makerchain}.
In other fields, several studies  have applied blockchain to maintain traceability~\cite{WANG2020103063,8718621}.
Those studies focused on the integrity of the information from another party.
Fabchain focuses on a single party's internal process to record each manufacturing process in a single party.
With the combination of Fabchain and Makerchain, we might verify that a party has manufactured a subset of a product and the party's manufacturing process.

3D printing technology is one of the most significant technologies associated with Industry 4.0.
Within the field of 3D printing, several studies have applied blockchain technology to guarantee the copyrights of 3D models~\cite{IPwithbc,belikovetsky20193d,mcconaghy2017visibility}.
In Industry 4.0, Fabchain could be the underlying scheme for those use cases.

Other studies have utilized blockchain technology in supply chain management and the detection of counterfeit products~\cite{poms,kennedy2017enhanced,maroun2019adoption}.
Combined with the ideas proposed in the studies described above, Fabchain would ensure transparency in the fabrication and distribution processes.

\section{Conclusion}
\label{Conclusion}
In this paper, we proposed and implemented Fabchain, which manages a 3D print job over a blockchain.
To clarify who is liable for the product, a fabricator should keep fabrication history in an immutable, and sustainably accessible manner without a trusted third party.
Fabchain can produce an immutable and publicly accessible fabrication history, due to its scheme that utilizes blockchain as an audit-able communication channel.
We implemented a prototype of Fabchain and performed experiments on the Ropsten network.
We concluded that our prototype can manage communication of a print job sequence within less than one minute so that Fabchain was reasonable for 3D printing.
In our evaluation scenario, Fabchain can satisfy the requirements of immutability and accessibility.
Finally, we described some open issues about the identification of entities and volatility of cost, and applicability of Fabchain.
In future work, we plan to implement applications which enable everyone to fabricate products with liability.

\bibliographystyle{IEEEtran}
\bibliography{myBibTex}
\end{document}